\documentclass[10pt,letterpaper,twocolumn]{article} 

\usepackage{ol2}
\usepackage[draft]{hyperref}
\usepackage{amsmath}

\begin{document}

\twocolumn[ 

\title{Transfer matrix for treating stratified media including birefringent crystals}


\author{Thomas Essinger-Hileman,$^{1,2}$}

\address{
$^1$The Johns Hopkins University, Department of Physics and Astronomy, 3400 N. Charles St., Baltimore, MD 21218
$^2$Princeton University, Department of Physics, Jadwin Hall, Princeton, NJ 08544
\\
}

\begin{abstract}
Birefringent crystals are extensively used to manipulate polarized light. The generalized transfer matrix developed allows efficient calculation of the full polarization state of light transmitted through and reflected by a stack of arbitrarily-many discrete layers of isotropic and birefringent materials at any frequency and angle of incidence. The matrix of a uniaxial birefringent crystal with arbitrary rotation is calculated, along with its reduction to the matrix of an isotropic medium. This method is of great practical importance where tight control of systematic effects is needed in optical systems employing birefringent crystals, one example being wave plates used by cosmic microwave background polarimetry with wide field-of-view telescopes. 
\end{abstract}


 ] 

\noindent Birefringent crystals are used to control and modify polarized light in a wide variety of applications from the millimeter-wave to the ultraviolet. Jones- or Mueller-matrix formalisms, which couple incoming and outgoing polarization states represented as vectors, offer useful characterizations of such devices. Idealized Jones and Mueller matrices for most devices are found in standard textbooks on the subject \cite{GoldsteinPolarizedLight, 1996aspo.book.....T, 1990plos.book.....K}. These ideal matrices are adequate in many applications; however, a general method for computing Jones and Mueller matrices for non-ideal devices including the effects of multiple reflections between devices is important where high precision and tight control of systematic errors is necessary. 

Real devices significantly diverge from ideal behavior when used across a wide frequency band or a range of angles of incidence. In addition, anti-reflection coatings that are important in creating high-throughput, birefringent-crystal devices are themselves sources of non-ideal behavior. And often wave plates are made of multiple layers of birefringent materials oriented relative to one another in such a way as to increase the frequency bandwidth \cite{1955PancharatnamAchromaticWaveplate}. 

Matrix methods for treating stratified media have a long history. Abel\`{e}s developed a $2 \times 2$ characteristic matrix method for fast computation of transmission and reflection for stratified isotropic media \cite{1950AbelesStratifiedMatrixMethod}. In Abel\`{e}s' treatment separate characteristic matrices were required for s- and p-polarized plane waves. See also \cite{1999prop.book.....B, 1987opt2.book.....H} for treatments of isotropic stratified media. A number of authors have expanded to using $4 \times 4$ matrices to treat birefringent materials that introduce mixing between orthogonal polarizations. Berreman and others developed matrix methods written in terms of first-order differential equations \cite{1972JOSA...62..502B}. Yeh derived a matrix method coupling propagating electromagnetic modes \cite{1979JOSA...69..742Y}.

We present an alternative derivation of a $4 \times 4$ matrix that directly couples total-field components at the interfaces between layers in a stratified, possibly birefringent, medium. The matrix method developed allows for full calculation of the polarization state of both transmitted and reflected waves at any frequency and angle of incidence of a stack of any number of isotropic and birefringent crystals, assumed to vary only in the z direction and be infinite in the x- and y-directions.

This treatment allows direct connection with the Jones and Mueller matrices, as shown below. Explicit formulas for a case of great practical importance, the matrix of a uniaxial crystal with its optic axis in the x-y plane, are derived. These results were first published in the author's doctoral dissertation \cite{EssingerHilemanPrincetonThesis} with particular application to millimeter-wave polarimetry.

\section{Parametrizations of Polarization}
The polarization state of a plane electromagnetic wave can be represented in multiple equivalent ways. The two-element Jones vector, $( |E_{x}| \exp(\imath \delta_{x}), |E_{y}| \exp(\imath \delta_{y}))$, gives the amplitude and phase of two orthogonal, time-harmonic electric fields as complex numbers. One can alternatively parametrize the polarization state of a plane wave using the Stokes parameters, defined as 

\begin{equation}
P = \langle \boldsymbol{E} \boldsymbol{E}^{\dagger} \rangle = I \sigma_{I} + Q \sigma_{Q} + U \sigma_{U} + V \sigma_{V},
\label{eqn:stokes_pol_matrix}
\end{equation}

\noindent where the $\sigma_{\imath}$ are the Pauli matrices\footnote{Note that this ordering of the Pauli matrices is different from that often used in the literature. \begin{equation}\begin{array}{cc}
	\sigma_{I} = \left(\begin{array}{cc}
		1 & 0 \\
		0 & 1 \\
		\end{array}\right) &
	\sigma_{Q} = \left(\begin{array}{cc}
		1 & 0 \\
		0 & -1 \\
		\end{array}\right) \\ \\
	\sigma_{U} = \left(\begin{array}{cc}
		0 & 1 \\
		1 & 0 \\
		\end{array}\right) &
	\sigma_{V} = \left(\begin{array}{cc}
		0 & -\imath \\
		\imath & 0 \\
		\end{array}\right) 
	\end{array}
\end{equation}}, $\boldsymbol{E}$ is the complex electric field vector, $\boldsymbol{E}^{\dagger}$ is its complex conjugate, and angled brackets represent averaging over a time that is long compared with the period of the electromagnetic wave, but short compared with time scales of the measurement. The four Stokes parameters can be gathered into a vector $(I, Q, U, V)$ with $4 \times 4$, real-valued Mueller matrices relating incoming and outgoing Stokes vectors. Given a Jones matrix, the expression

\begin{equation}
M_{ij} = \frac{1}{2} Tr(\sigma_{i} J \sigma_{j} J^{\dagger}),
\label{eqn:jones_mueller_relation}
\end{equation}

\noindent where $\imath$ and $\jmath$ are in the set $\{ I, Q, U, V \}$, can be used to calculate the corresponding Mueller matrix \cite{2007A&A...470..771J}. It should be noted that Mueller matrices can be band-averaged element-by-element \cite{2010ApOpt..49.6313B} making the matrix method described applicable to experiments with finite bandwidth in a straightforward way.

\section{The generalized transfer matrix}

We seek to relate the total, tangential, complex $\boldsymbol{E}$ and $\boldsymbol{H}$ field components at the two interfaces of a system that is infinite in the x and y directions and varies only in the z direction. The plane of incidence is assumed to be the x-z plane. We will connect the vector $\Lambda = \left( E^{(x)}, H^{(y)}, E^{(y)}, -H^{(x)} \right)$ on the two interfaces. The great advantage of the generalized transfer matrix is that it directly relates the field amplitudes at the two interfaces, taking into account multiple reflections and allowing the optical properties of a system involving arbitrarily-many layers to be calculated by simple matrix multiplication. 

\begin{figure*}[htb]
\centerline{\includegraphics[width=13cm]{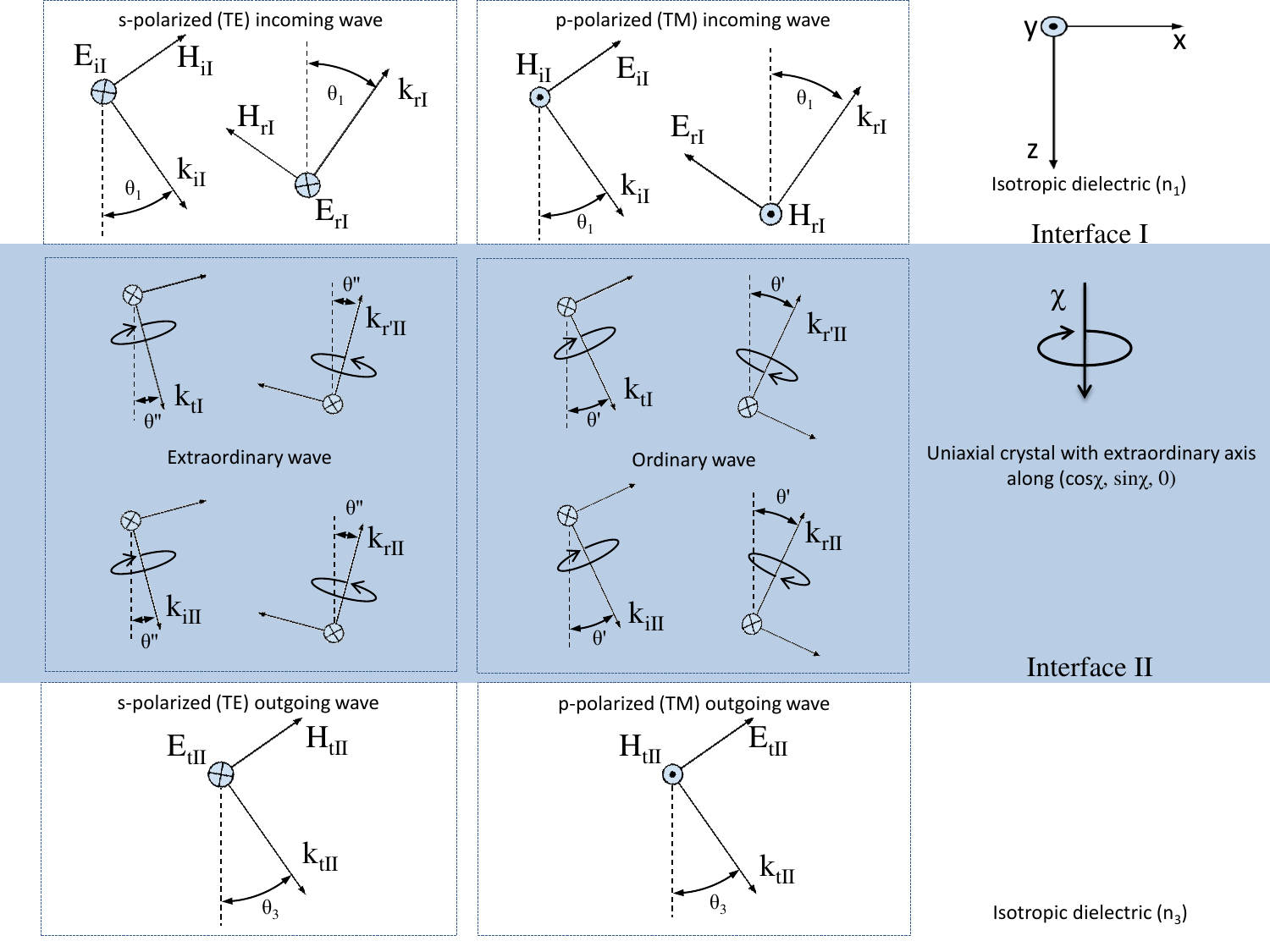}}
\caption{Geometry of the rays used to calculate the generalized transfer matrix for the uniaxial crystal. The two polarization states defined by the plane of incidence, the s- and p-waves, are mixed inside the uniaxial crystal into the ordinary and extraordinary waves.}
\label{fig:fields}
\end{figure*}

Once the generalized transfer matrix is computed for a system, it is straightforward to calculate the amplitude transmission, reflection, and absorption coefficients for a wave incident upon the system from one isotropic dielectric of index $n_{1}$ and transmitted through to another isotropic dielectric of index $n_{3}$. This situation is depicted in Figure \ref{fig:fields}, where the specific case of a birefringent crystal is shown; however, we are presently only concerned with the fields on the outsides of the crystal. For our present purposes, one could replace the anisotropic crystal with any system of isotropic and birefringent crystals represented by a total generalized transfer matrix, T. If we denote reflected and transmitted waves by subscripts r and t, respectively, and s- and p-polarizations with superscripts, the fields at the interfaces are related via the generalized transfer matrix as

\begin{equation} 
	\left( \begin{array}{c} 
		( E^{p}_{i} + E^{p}_{r} ) \cos \theta_{1} \\ n_{1} (E^{p}_{i} - E^{p}_{r}) \\ ( E^{s}_{i} + E^{s}_{r}) \\ (E^{s}_{i} - E^{s}_{r}) n_{1} \cos \theta_{1}
	\end{array} \right) = 	
T	\left( \begin{array}{c} 
		E^{p}_{t} \cos \theta_{3} \\ n_{3} E^{p}_{t} \\ E^{s}_{t} \\ E^{s}_{t} n_{3} \cos \theta_{3}
	\end{array} \right).
\label{eqn:r_t_amplitude_conditions}
\end{equation}

The plus and minus signs between incident and reflected magnetic field components on the left-hand side of this equation are dictated by the boundary conditions. When the z component of $\boldsymbol{k}$ changes sign for the incident versus reflected waves, the components of $\boldsymbol{E}$ and $\boldsymbol{H}$ follow suit according to $\boldsymbol{H} = n \boldsymbol{k} \times \boldsymbol{E}$. The only way to simultaneously satisfy the boundary condition in the fields is with a 180$^{\circ}$ phase shift in those components. 

\subsection{Jones and Mueller matrices}
The above system of equations can be solved to give E$^{p}_{r}$, E$^{s}_{r}$, E$^{p}_{t}$, and E$^{s}_{t}$ in terms of E$^{p}_{i}$ and E$^{s}_{i}$. Our goal is to write these relations in the form of Jones matrices for transmission and reflection, which can then be translated to Mueller matrices using Equation \ref{eqn:jones_mueller_relation}. Denoting the components of the 4 x 4 generalized transfer matrix as $t_{\imath \jmath}$, we can simplify our expressions by defining

\begin{equation} \begin{array}{rcl}
	\alpha & = & (t_{11} \cos \theta_{3} + t_{12} n_{3}) / \cos \theta_{1} \\
	\beta  & = & (t_{13} + t_{14} n_{3} \cos \theta_{3}) /  \cos \theta_{1} \\
	\gamma & = & (t_{21}  \cos \theta_{3} + t_{22} n_{3} ) / n_{1} \\
	\delta & = & (t_{23} + t_{24} n_{3} \cos \theta_{3} ) / n_{1} \\
	\eta   & = & ( t_{31}  \cos \theta_{3} + t_{32} n_{3})\\
	\kappa & = & (t_{33}  + t_{34} n_{3} \cos \theta_{3}) \\
	\rho   & = & ( t_{41}  \cos \theta_{3} + t_{42} n_{3}) / (n_{1} \cos \theta_{1}) \\
	\sigma & = & ( t_{43}  + t_{44} n_{3} \cos \theta_{3}) / (n_{1} \cos \theta_{1}) \\
  \Gamma & = & [(\alpha + \gamma) (\kappa + \sigma) - (\beta + \delta) (\eta + \rho)]^{-1} .\\
	\end{array}
\end{equation}

\noindent In terms of these constants, the transmitted-wave amplitudes are 

\begin{equation}\begin{array}{rcl}
	\left( \begin{array}{c} E^{p}_{t} \\ E^{s}_{t} \end{array} \right)
	& = &
	\left( \begin{array}{cc} 
		J^{t}_{11} & J^{t}_{12} \\
		J^{t}_{21} & J^{t}_{22} \\
		\end{array} \right)
	\left( \begin{array}{c} E^{p}_{i} \\ E^{s}_{i} \end{array} \right) \\
	& = &
	2 \Gamma
	\left( \begin{array}{cc} 
		\kappa + \sigma & -\beta - \delta \\
		-\eta - \rho    & \alpha + \gamma \\
		\end{array} \right)
	\left( \begin{array}{c} E^{p}_{i} \\ E^{s}_{i} \end{array} \right).
\end{array}
\end{equation}

\noindent The reflected-wave amplitudes are

\begin{equation}
	\left( \begin{array}{c} E^{p}_{r} \\ E^{s}_{r} \end{array} \right) =	\left( \begin{array}{cc} 
			J^{r}_{11} & J^{r}_{12} \\
			J^{r}_{21} & J^{r}_{22} \\
			\end{array} \right) \left( \begin{array}{c} 
				E^{p}_{i} \\ E^{s}_{i} \end{array} \right), 
\end{equation}

\noindent with

\begin{equation} \begin{array}{rcl}
	J^{r}_{11} & = & \Gamma ((\gamma - \alpha)(\kappa + \sigma) - (\delta - \beta)(\eta + \rho)) \\
	J^{r}_{12} & = & 2 \Gamma (\alpha \delta - \gamma \beta) \\
	J^{r}_{21} & = & 2 \Gamma (\eta \sigma - \rho \kappa ) \\
	J^{r}_{22} & = & \Gamma ((\alpha + \gamma)(\kappa - \sigma) - (\beta + \delta)(\eta - \rho)). \\
	\end{array}
\end{equation}

The Jones matrices for the reflected and transmitted waves can then be transformed into Mueller matrices using Equation \ref{eqn:jones_mueller_relation}. 

\section{Matrix of a uniaxial crystal}
We now derive the transfer matrix for a uniaxial, birefringent crystal with its optic axis parallel to the faces of the crystal in the x-y plane. For full treatments of wave propagation in birefringent media and refraction at an interface between birefringent media see \cite{1999prop.book.....B, 1983ChenCoordinateFreeApproach, 1928HandbuchderPhysikSzivessy}.

\subsection{Electromagnetic waves in a uniaxial crystal}
Anisotropic, non-magnetic, linear dielectrics can be characterized via their dielectric tensor, $\epsilon_{kl}$, which transforms the electric field vector in the material to the electric displacement. In the special case considered here of a uniaxial crystal with its optic axis tangent to interfaces I and II and rotated at an arbitrary angle, $\chi$, to the x-axis,

\begin{equation}
\varepsilon^{\prime} =	R \left( \chi \right)	\left( \begin{array}{ccc} 
		n_{e}^{2} & 0 & 0 \\
		0 & n_{o}^{2} & 0 \\
		0 & 0 & n_{o}^{2}
	\end{array} \right) R \left( - \chi \right) ,
\label{eqn:rotated_dielectric_tensor1}
\end{equation}

\noindent where $R(\chi)$ is the rotation matrix by $\chi$ about the z axis

\begin{equation}
R(\chi) = \left(\begin{array}{ccc} 
		\cos \chi & -\sin \chi & 0 \\
		\sin \chi & \cos \chi & 0 \\
		0 & 0 & 1
	\end{array} \right)
\label{eqn:rotated_dielectric_tensor2}
\end{equation}

\noindent The above matrix holds for a uniaxial crystal with its optic axis pointed in the direction of $\hat{\boldsymbol{\xi}} = (\cos \chi, \sin \chi, 0)$. The crystal is assumed to have ordinary and extraordinary indices of refraction of $n_{o}$ and $n_{e}$. The inverse to $\varepsilon^{\prime}$ transforms $\boldsymbol{D}$ to $\boldsymbol{E}$.

One consequence of the anisotropy of the crystal is that the electric field vector, $\boldsymbol{E}$, and the electric displacement, $\boldsymbol{D} = \varepsilon \boldsymbol{E}$, of an EM wave no longer point in the same direction. Both $\boldsymbol{E}$ and $\boldsymbol{D}$ remain perpendicular to the magnetic field, $\boldsymbol{H}$, and the direction of energy transport is still given by the Poynting Vector, $\boldsymbol{S} = \boldsymbol{E} \times \boldsymbol{H}$; however, the wave propagates in the direction of $\boldsymbol{D} \times \boldsymbol{H}$, given by the normal direction to planes of constant phase. It is therefore the triplet of vectors $\hat{\boldsymbol{k}}$, $\boldsymbol{D}$, and $\boldsymbol{H}$ that are mutually orthogonal. This makes it useful to propagate components of $\boldsymbol{D}$ and then transform to $\boldsymbol{E}$ to match boundary conditions at the interfaces. This is especially true since the magnitudes of $\boldsymbol{D}$ and $\boldsymbol{H}$ are related simply by

\begin{equation}
| \boldsymbol{H} | = \frac{1}{n} | \boldsymbol{D} |.
\label{eqn:H_D_magnitudes}
\end{equation}

Given $\hat{\boldsymbol{k}}$ and either $\boldsymbol{E}$ or $\boldsymbol{D}$, the other vectors for the wave can be calculated through Equations \ref{eqn:rotated_dielectric_tensor1} and \ref{eqn:rotated_dielectric_tensor2}, as well as the fact that

\begin{equation}
	\boldsymbol{H} = \frac{1}{n} \hat{\boldsymbol{k}} \times \boldsymbol{D} .
\label{eqn:uniaxial_H_D_relation}
\end{equation}
 
\noindent Here n is a real-valued, angle-dependent refractive index which gives the speed of a ray as it traverses the medium, v$_{p} = c / n$, that will be calculated below. 

Another consequence of the anisotropy of the material is that a linearly-polarized plane wave at a dielectric boundary gets refracted into two separate plane waves polarized along preferred directions in the crystal unless the incoming plane wave happens to be polarized along one of the principle axes of the crystal. These two waves travel at different speeds. For a uniaxial crystal, one of these waves, the ordinary ray, always travels at the same speed, given by

\begin{equation}
v^{\prime} = v_{o} = c / n_{o},
\label{eqn:ordinary_ray_velocity}
\end{equation}

\noindent while the extraordinary ray has a velocity which depends upon the angle $\psi$ between the wave and the optical axis and is given by

\begin{equation}\begin{array}{ccccc}
v^{\prime\prime} & = & \frac{c}{n^{\prime\prime}} & = & \left(v_{o}^{2} \cos^{2} \psi + v_{e}^{2} \sin^{2} \psi \right)^{1/2} \\
& & & = & c \left(\frac{\cos^{2} \psi}{n^{2}_{o}} + \frac{\sin^{2} \psi}{n^{2}_{e}}  \right)^{1/2} \end{array}.
\label{eqn:extraordinary_ray_velocity}
\end{equation}

An EM wave which is incident at angle $\theta_{1}$ upon a surface of the uniaxial crystal from an isotropic dielectric of index $n_{1}$ will then refract by two separate angles, which are each individually given by the familiar Snell's Law

\begin{equation}
n_{o} \sin \theta^{\prime} = n^{\prime\prime} \sin \theta^{\prime\prime} = n_{1} \sin \theta_{1}.
\label{eqn:uniaxial_eo_snell}
\end{equation}

\noindent The two rays remain in the plane of incidence. The ordinary ray is always at the same angle relative to the angle of incidence. The equation for the extraordinary ray is complicated by the fact that the index of refraction depends on the angle through the material, which depends on the angle of incidence and orientation of the crystal, $\chi$, in a nontrivial manner. Let the faces of the uniaxial crystal lie in the x-y plane. The extraordinary ray that propagates through the material at angle $\theta,^{\prime\prime}$ yet to be determined, has unit propagation vector $\hat{\boldsymbol{k}}^{\prime\prime} = \sin \theta^{\prime\prime} \hat{\boldsymbol{x}} + \cos \theta^{\prime\prime} \hat{\boldsymbol{z}}$. The angle $\psi$, defined in Equation \ref{eqn:extraordinary_ray_velocity}, between these two is then $\cos \psi = \hat{\boldsymbol{k}^{\prime\prime}} \cdot \hat{\boldsymbol{\xi}} = \sin \theta^{\prime\prime} \cos \chi$. Thus Equation \ref{eqn:extraordinary_ray_velocity} becomes

\begin{equation} \begin{array}{rcl}
(1/n^{\prime\prime})^{2} & = & \cos \psi/n_{o}^{2} + \sin \psi / n_{e}^{2} \\
  & = & (1/n_{e})^{2} + (n_{o}^{-2}-n_{e}^{-2})\sin^{2} \theta^{\prime\prime} \cos^{2} \chi \\
	\end{array}.
\label{eqn:extraordinary_ray_velocity2}
\end{equation}

\noindent Using Equation \ref{eqn:uniaxial_eo_snell} to eliminate $\sin \theta^{\prime\prime}$ and solving for $n^{\prime\prime}$ yields

\begin{equation}
n^{\prime\prime} = n_{e} \sqrt{ 1 + (n_{e}^{-2}-n_{o}^{-2}) n_{1}^{2} \sin^{2} \theta_{1} \cos^{2} \chi } \hspace{0.1in},
\label{eqn:extraordinary_index}
\end{equation}

\noindent and $\sin\theta^{\prime\prime}=(n_{1}/ n^{\prime\prime})\sin\theta_{1}$.

Equation \ref{eqn:uniaxial_eo_snell}, together with Equation \ref{eqn:extraordinary_index}, gives the angles of the ordinary and extraordinary rays for a given angle of incidence and rotation of the HWP. These rays both travel in the x-z plane and have associated unit-normal vectors $\hat{\boldsymbol{k}}^{\prime}$ and $\hat{\boldsymbol{k}}^{\prime\prime}$, which are 

\begin{equation} \begin{array}{lr}
\hat{\boldsymbol{k}}^{\prime} = ( \sin \theta^{\prime}, 0, \cos \theta^{\prime}); &
\hat{\boldsymbol{k}}^{\prime\prime} = ( \sin \theta^{\prime\prime}, 0, \cos \theta^{\prime\prime}) \\
\end{array}.
\label{eqn:eo_unit_normals}
\end{equation}

The directions of vibration of the electric displacement vector, or in other words the polarizations of the two refracted rays, can also be calculated. The ordinary ray, traveling through the material at angle $\theta^{\prime}$, must have its electric displacement perpendicular to both the direction of propagation $\hat{\boldsymbol{k}}^{\prime}$ and the direction of the optical axis $\hat{\boldsymbol{\xi}}$. Thus

\begin{equation}
\hat{D}^{\prime} = \frac{\boldsymbol{D}^{\prime}}{| \boldsymbol{D^{\prime}} |} = \alpha^{\prime} \hat{\boldsymbol{k}}^{\prime} \times \hat{\boldsymbol{\xi}} = \alpha^{\prime} \left( \begin{array}{c} - \sin \chi \cos \theta^{\prime} \\ \cos \chi \cos \theta^{\prime} \\ \sin \chi \sin \theta^{\prime} \\  \end{array} \right)
\label{eqn:uniaxial_D_ordinary}
\end{equation}

\noindent Similarly, the extraordinary ray must have an electric displacement that is perpendicular to its direction of propagation and to the direction of vibration of the ordinary ray. It is thus

\begin{equation}\begin{array}{rcl}
\hat{\boldsymbol{D}}^{\prime\prime} & = & \boldsymbol{D}^{\prime\prime} / | \boldsymbol{D^{\prime\prime}} | = \alpha^{\prime\prime} \hat{\boldsymbol{k}}^{\prime\prime} \times \hat{\boldsymbol{D}}^{\prime} \\
 & & \\
 & = & \alpha^{\prime\prime} 
\left( \begin{array}{c} 
- \cos \chi \cos \theta^{\prime} \cos \theta^{\prime\prime} \\ 
- \sin \chi \left[ \sin \theta^{\prime} \sin \theta^{\prime\prime} + \cos \theta^{\prime} \cos \theta^{\prime\prime} \right] \\
\cos \chi \cos \theta^{\prime} \sin \theta^{\prime\prime} \\ 
\end{array} \right) \end{array}
\label{eqn:uniaxial_D_extraordinary}
\end{equation}

\noindent The constants $\alpha^{\prime}$ and $\alpha^{\prime\prime}$ normalize these vectors.

\subsection{Relations between field components}
The geometry under consideration is shown in Figure \ref{fig:fields}. A plane wave of frequency $\nu$ is incident on a birefringent dielectric. The plane of incidence is the x-z plane. We wish to derive relationships between tangential $\boldsymbol{E}$ and $\boldsymbol{H}$ at both of the interfaces, labeled I and II. In the absence of free charges on the boundary, the tangential components of both $\boldsymbol{E}$ and $\boldsymbol{H}$ are continuous. We will thus consider the relationship between the vector (E$_{x}$, H$_{y}$, E$_{y}$, -H$_{x}$) on the two boundaries; however, we will propagate components of $\boldsymbol{D}$ because it is perpendicular to $\hat{\boldsymbol{k}}$.

In the calculations below, subscripts will denote incoming, transmitted, and reflected field amplitudes as lower-case i, t, and r, respectively, along with a roman-numeral I or II denoting the interface concerned. Subscripts s and p will denote waves with electric field vector perpendicular to the plane of incidence (s-polarized wave) and in the plane of incidence (p-polarized wave) in the surrounding isotropic medium. In the uniaxial crystal $^{\prime}$ or $^{\prime\prime}$ will denote the ordinary or extraordinary rays. As an example, $E^{\prime\prime}_{tI}$ is the electric field amplitude of the extraordinary ray that is transmitted from interface I. A subscript of ``r$^{\prime}$II'' denotes the ray reflected off interface II which has traversed the medium and is incident from below on interface I.

Rays that traverse the uniaxial crystal develop a phase shift that depends on their angle, the thickness of the crystal, and the refractive index seen by that ray. The phase shifts differ for the ordinary and extraordinary rays, and are given by 

\begin{equation} \begin{array}{rcl}
	\delta^{\prime}       & = & \tilde{n}_{o} t \cos \theta^{\prime} \\
	\delta^{\prime\prime} & = & \tilde{n}^{\prime\prime} t \cos \theta^{\prime\prime}
\end{array},
\label{eqn:sapphire_model_phase_shift}
\end{equation}

\noindent where $\tilde{n} = n (1 - \imath \tan \delta)^{1/2}$ is the complex refractive index, the imaginary part of which is often characterized by the loss tangent, $\tan \delta$, and encodes dielectric loss in the material. The real part is the refractive index given above. Rays that are transmitted from Interface I are incident on Interface II with a phase shift. This relates $D_{tI}$ and $D_{iII}$ as

\begin{equation} \begin{array}{rcl}
	D^{\prime}_{iII} & = & D^{\prime}_{tI} \exp(\imath k_{0} \delta^{\prime}) \\
	D^{\prime\prime}_{iII} & = & D^{\prime\prime}_{tI} \exp(\imath k_{0} \delta^{\prime\prime})
\end{array}.
\label{eqn:sapphire_model_propagation1}
\end{equation}

\noindent Likewise, the ray that reflects off interface II is phase shifted on its way to interface I, giving

\begin{equation} \begin{array}{rcl}
	D^{\prime}_{rII} & = & D^{\prime}_{r^{\prime}II} \exp(- \imath k_{0} \delta^{\prime}) \\
	D^{\prime\prime}_{rII} & = & D^{\prime\prime}_{r^{\prime}II} \exp(- \imath k_{0} \delta^{\prime\prime})
\end{array}.
\label{eqn:sapphire_model_propagation2}
\end{equation}

In the above, $k_{0}$ is the wave number for the wave in vacuum, equal to $2 \pi / \lambda_{0}$.  For concision, we can break $\boldsymbol{D}$ and $\boldsymbol{H}$ into unknown total magnitudes multiplied by unit-vector directions that are known from Equations \ref{eqn:uniaxial_D_ordinary}, \ref{eqn:uniaxial_D_extraordinary}, and \ref{eqn:uniaxial_H_D_relation}. The components of $\boldsymbol{D}$ and $\boldsymbol{H}$ transmitted from Interface I can be written out explicitly as

\begin{equation} \begin{array}{rcl}\vspace{3pt}
	\boldsymbol{D}^{\prime}_{tI}                               & = & | \boldsymbol{D}^{\prime}_{tI} | (\hat{D}^{\prime \text{(x)}}_{tI},\hspace{3pt} \hat{D}^{\prime \text{(y)}}_{tI},\hspace{3pt} \hat{D}^{\prime \text{(z)}}_{tI}) \\ \vspace{3pt}
	\boldsymbol{D}^{\prime\prime}_{tI}                         & = & | \boldsymbol{D}^{\prime\prime}_{tI} | (\hat{D}^{\prime\prime \text{(x)}}_{tI},\hspace{3pt} \hat{D}^{\prime\prime \text{(y)}}_{tI},\hspace{3pt} \hat{D}^{\prime\prime \text{(z)}}_{tI}) \\ \vspace{3pt}
	\boldsymbol{H}^{\prime}_{tI}                               & = & \frac{1}{n^{\prime}} | \boldsymbol{D}^{\prime}_{tI} | (\hat{H}^{\prime \text{(x)}}_{tI},\hspace{3pt} \hat{H}^{\prime \text{(y)}}_{tI},\hspace{3pt} \hat{H}^{\prime \text{(z)}}_{tI}) \\ \vspace{3pt}
	\boldsymbol{H}^{\prime\prime}_{tI}                         & = & \frac{1}{n^{\prime\prime}} | \boldsymbol{D}^{\prime\prime}_{tI} | (\hat{H}^{\prime\prime \text{(x)}}_{tI},\hspace{3pt} \hat{H}^{\prime\prime \text{(y)}}_{tI},\hspace{3pt} \hat{H}^{\prime\prime \text{(z)}}_{tI})
	\end{array},
\label{eqn:bc_abbreviations1}
\end{equation}

\noindent where use has been made of the relationship between the magnitudes of $\boldsymbol{D}$ and $\boldsymbol{H}$, Equation \ref{eqn:H_D_magnitudes}. The complex components of the field unit vectors, such as $D^{\prime \text{(x)}}_{tI}$, are known and given in explicit form in Equation \ref{eqn:tabulated_field_components} below. All other vectors at the two interfaces, shown in Figure \ref{fig:fields}, can be written in terms of the known transmitted field directions from interface I, as well as undetermined field amplitudes. The remaining field components at interface I can be written 

\begin{equation} \begin{array}{rcl}\vspace{3pt}
	\boldsymbol{D}^{\prime}_{\text{r}^{\prime}\text{II}}       & = & |\boldsymbol{D}^{\prime}_{\text{r}^{\prime}\text{II}}| (\hat{D}^{\prime \text{(x)}}_{tI},\hspace{3pt} \hat{D}^{\prime \text{(y)}}_{tI},\hspace{3pt} -\hat{D}^{\prime \text{(z)}}_{tI}) \\ \vspace{3pt}
	\boldsymbol{D}^{\prime\prime}_{\text{r}^{\prime}\text{II}} & = & |\boldsymbol{D}^{\prime\prime}_{\text{r}^{\prime}\text{II}}| (\hat{D}^{\prime\prime \text{(x)}}_{tI},\hspace{3pt} \hat{D}^{\prime\prime \text{(y)}}_{tI},\hspace{3pt} -\hat{D}^{\prime\prime \text{(z)}}_{tI})\\ \vspace{3pt}
	\boldsymbol{H}^{\prime}_{\text{r}^{\prime}\text{II}}       & = & \frac{1}{n^{\prime}} |\boldsymbol{D}^{\prime}_{\text{r}^{\prime}\text{II}}| (-\hat{H}^{\prime \text{(x)}}_{tI},\hspace{3pt} -\hat{H}^{\prime \text{(y)}}_{tI},\hspace{3pt} \hat{H}^{\prime \text{(z)}}_{tI}) \\ \vspace{3pt}
	\boldsymbol{H}^{\prime\prime}_{\text{r}^{\prime}\text{II}} & = & \frac{1}{n^{\prime\prime}} |\boldsymbol{D}^{\prime\prime}_{\text{r}^{\prime}\text{II}}| (-\hat{H}^{\prime\prime \text{(x)}}_{tI},\hspace{3pt} -\hat{H}^{\prime\prime \text{(y)}}_{tI},\hspace{3pt} 
	\hat{H}^{\prime\prime \text{(z)}}_{tI})
	\end{array}.
\label{eqn:bc_abbreviations2}
\end{equation}

\noindent Because the wave vectors of the reflected waves flip signs in the z-direction, the z-components of $\boldsymbol{D}_{r}$ and the x- and y-components of $\boldsymbol{H}_{r}$ must change signs relative to Equation \ref{eqn:bc_abbreviations1} to satisfy Equation \ref{eqn:uniaxial_H_D_relation}.

A system of four equations relating the total-field x- and y-components at Interface I and the individual ray components (tI and r$^{\prime}$II) can be written in matrix form as $\Lambda_{I} =(E^{x}_{I}, H^{y}_{I}, E^{y}_{I}, - H^{x}_{I}) = \Psi \Phi_{I} X$ with

\begin{equation}
	\Phi_{I} = 	
	\left( \begin{array}{cccc}
		\hat{D}^{\prime \text{(x)}}_{tI} & \hat{D}^{\prime\prime \text{(x)}}_{tI} & \hat{D}^{\prime \text{(x)}}_{tI} & \hat{D}^{\prime\prime \text{(x)}}_{tI} \\
		\frac{1}{n^{\prime}} \hat{H}^{\prime \text{(y)}}_{tI} & \frac{1}{n^{\prime\prime}} \hat{H}^{\prime\prime \text{(y)}}_{tI} & -\frac{1}{n^{\prime}} \hat{H}^{\prime \text{(y)}}_{tI} & -\frac{1}{n^{\prime\prime}} \hat{H}^{\prime\prime \text{(y)}}_{tI} \\
		\hat{D}^{\prime \text{(y)}}_{tI} & \hat{D}^{\prime\prime \text{(y)}}_{tI} & \hat{D}^{\prime \text{(y)}}_{tI} & \hat{D}^{\prime\prime \text{(y)}}_{tI} \\
		-\frac{1}{n^{\prime}} \hat{H}^{\prime \text{(x)}}_{tI} & - \frac{1}{n^{\prime\prime}} \hat{H}^{\prime\prime \text{(x)}}_{tI} & \frac{1}{n^{\prime}} \hat{H}^{\prime \text{(x)}}_{tI} & \frac{1}{n^{\prime\prime}} \hat{H}^{\prime\prime \text{(x)}}_{tI} \\
	\end{array} \right),
\label{eqn:boundary_conditions_interfaceI}
\end{equation}

\noindent and $X = (|\boldsymbol{D}^{\prime}_{tI}|, |\boldsymbol{D}^{\prime\prime}_{tI}|, |\boldsymbol{D}^{\prime}_{r^{\prime}II}|, |\boldsymbol{D}^{\prime\prime}_{r^{\prime}II}|)$. The matrix $\Psi$ converts components of $\boldsymbol{D}$ to $\boldsymbol{E}$, and is given by

\begin{equation}
	\Psi_{I} = 	
	\left( \begin{array}{cccc}
		(\epsilon^{ -1})^{\prime}_{xx} & 0 &(\epsilon^{ -1})^{\prime}_{xy}& 0 \\
		0 & 1 & 0 & 0 \\
		(\epsilon^{ -1})^{\prime}_{yx} & 0 & (\epsilon^{ -1})^{\prime}_{yy} & 0 \\
		0 & 0 & 0 & 1 
	\end{array} \right),
\label{eqn:psi_matrix}
\end{equation}

\noindent in terms of components of $(\epsilon^{-1})^{\prime} = \left[ R(\chi) \epsilon R(-\chi) \right]_{,}^{-1}$ the inverse of the dielectric tensor for the rotated layer.

A similar relation holds for Interface II, where Equations \ref{eqn:sapphire_model_propagation1} and \ref{eqn:sapphire_model_propagation2} can be used to relate the fields on Interface II to those on Interface I. Again, the relation can be written in matrix form as $\Lambda_{II}=(E^{x}_{II}, H^{y}_{II}, E^{y}_{II}, - H^{x}_{II})= \Psi \Phi_{II} X$ with $\Phi_{II} = \Phi_{I} P$. The matrix P gives the effect on the fields of propagating through the material and is given by

\begin{equation}
	P = 	
	\left( \begin{array}{cccc}
	 \exp(- \Delta^{\prime}) & 0 & 0 & 0 \\
	 0 & \exp(- \Delta^{\prime\prime}) & 0 & 0 \\
	 0 & 0 & \exp(\Delta^{\prime}) & 0 \\
	 0 & 0 & 0 & \exp(\Delta^{\prime\prime}) \\	
	\end{array} \right)
\label{eqn:boundary_conditions_interfaceII}
\end{equation}

\noindent where $\Delta^{\prime} = \imath k_{0} \delta^{\prime}$ and $\Delta^{\prime\prime} = \imath k_{0} \delta^{\prime\prime}$. The phases $\delta^{\prime}$ and $\delta^{\prime\prime}$ are defined in Equation \ref{eqn:sapphire_model_phase_shift}.

We can find the relation we are seeking between $\Lambda_{I}$ and $\Lambda_{II}$ by solving for X in the above equations and setting these equal to one another 

\begin{equation}\begin{array}{c}
\Lambda_{I} = \Psi \Phi_{I} (\Psi \Phi_{I} P)^{-1} \Lambda_{II} = T \Lambda_{II} \\[5pt]
T = \Psi \Phi_{I} P^{-1} \Phi_{I}^{-1} \Psi^{-1}
\end{array}.
\label{eqn:final_generalized_matrix}
\end{equation}

\noindent The matrix T is the generalized transfer matrix for a uniaxial crystal with its optical axis at an angle $\chi$ to the x axis. It should be stressed that this matrix deals with \textit{total} electric and magnetic fields at the two boundaries of the crystal, which allows the matrix to take into account multiple reflections. Explicit formulas for the field components transmitted from Interface I are

\begin{equation} \begin{array}{rcl}
	\hat{D}^{\prime}_{tI} & = & (-\sin \chi \cos \theta^{\prime},\hspace{3pt} \cos \chi \cos \theta^{\prime},\hspace{3pt} \sin \chi \sin \theta^{\prime}) \hspace{3pt}/ \\
	& & [\cos^{2} \theta^{\prime} + \sin^{2} \theta^{\prime} \sin^{2} \chi]^{1/2} \\
	& & \\	
	\hat{D}^{\prime\prime}_{tI} & = & (\cos \chi \cos \theta^{\prime} \cos \theta^{\prime\prime},\\
	& &  \sin \chi (\sin \theta^{\prime} \sin \theta^{\prime\prime} + \cos \theta^{\prime} \cos \theta^{\prime\prime}), \\
	& & \hspace{3pt} - \cos \chi \cos \theta^{\prime} \sin \theta^{\prime\prime}) \hspace{3pt}/ \\  & & [\cos^{2} \chi \cos^{2} \theta^{\prime} + \sin^{2} \chi \cos^{2} (\theta^{\prime} - \theta^{\prime\prime}) ]^{1/2} \\ 
	& & \\
	\hat{H}^{\prime}_{tI} & = & (- \cos^{2} \theta^{\prime} \cos \chi,\hspace{3pt} - \sin \chi,\hspace{3pt} \cos \theta^{\prime} \sin \theta^{\prime} \cos \chi)\hspace{3pt}/ \\
	& & [\cos^{2} \theta^{\prime} \cos^{2} \chi + \sin^{2} \chi ]^{1/2} \\
	& & \\	
	\hat{H}^{\prime\prime}_{tI} & = & (- \cos (\theta^{\prime} - \theta^{\prime\prime}) \cos \theta^{\prime\prime} \sin \chi,\hspace{3pt} \cos \theta^{\prime} \cos \chi, \\
	& & \cos (\theta^{\prime} - \theta^{\prime\prime}) \sin \theta^{\prime\prime} \sin \chi)\hspace{3pt}/ \\
	& & [ \cos^{2} (\theta^{\prime}-\theta^{\prime\prime}) \sin^{2} \chi + \cos^{2} \theta^{\prime} \cos^{2} \chi ]^{1/2}\\
\end{array}
\label{eqn:tabulated_field_components}
\end{equation}

\subsection{Matrix of an isotropic medium}
The above equations simplify considerably for the case of an isotropic dielectric. Specifically, s- and p-polarized waves no longer mix within the dielectric and the generalized transfer matrix becomes block diagonal. Taking $\chi$ to be zero, $\theta_{1}=\theta_{2}=\theta$, and $n^{\prime}=n^{\prime\prime}=n$, Equations \ref{eqn:tabulated_field_components} reduce to give $\hat{D}^{\prime}_{tI} = (0, 1, 0)$; $\hat{D}^{\prime\prime}_{tI} = (\cos \theta, 0, -\sin \theta)$; $\hat{H}^{\prime}_{tI} = (-\cos \theta, 0, \sin \theta)$; and $\hat{H}^{\prime\prime}_{tI} = (0, 1, 0)$. Carrying these field components through Equations \ref{eqn:boundary_conditions_interfaceI}, \ref{eqn:boundary_conditions_interfaceII}, and \ref{eqn:final_generalized_matrix} yields the generalized transfer matrix for an isotropic medium

\begin{equation}
	\left( \begin{array}{cccc}
		\cos k_{0} \delta & \frac{\imath}{n} \sin k_{0} \delta \cos \theta  & 0 & 0 \\
		\imath n \frac{\sin k_{0} \delta}{\cos \theta} & \cos k_{0} \delta & 0 & 0 \\
		0 & 0 & \cos k_{0} \delta & \imath \frac{\sin k_{0} \delta}{n \cos \theta} \\
		0 & 0 & \imath n \sin k_{0} \delta \cos \theta & \cos k_{0} \delta \\
	\end{array} \right).
\label{eqn:boundary_conditions_interfaceI}
\end{equation}

\noindent The $2 \times 2$ matrices in the upper left and lower right of the isotropic generalized transfer matrix are those familiar from the literature for s- and p-polarized light.

\section{Conclusion}
We have developed a matrix method capable of exact treatment of stratified optical systems involving birefringent crystals. This method allows straightforward calculation of transmitted and reflected polarization states given an incident polarization state, for any frequency and angle of incidence. The generalized transfer matrices of individual layers can be multiplied together to give the matrix of a system composed of arbitrarily-many layers. Because the generalized transfer matrix relates the total electric and magnetic fields at the two boundaries, multiple reflections between interfaces are automatically included in the treatment.

Though the generalized transfer matrix method is able to treat general birefringent crystals, we have limited ourselves to calculating the important case of a uniaxial crystal with its optic axis in the x-y plane. Full expressions for this case have been given, as well as the reduction of this result to isotropic media.

\vspace{8pt}
We express our sincere thanks to Lyman Page and Suzanne Staggs of Princeton University for many helpful discussions. This work was supported by the US National Science Foundation through awards AST-0408698 and PHY-0355328, as well as a National Defense Science and Engineering Graduate (NDSEG) Fellowship.

\bibliographystyle{osajnl}
\bibliography{researchBib}

\begin{thebibliography}{10}
\newcommand{\enquote}[1]{``#1''}

\bibitem{GoldsteinPolarizedLight}
D.~H. Goldstein, \emph{{Polarized Light}} ({CRC Press, Boca Raton, FL}, 2011).

\bibitem{1996aspo.book.....T}
J.~{Tinbergen}, \emph{{Astronomical Polarimetry}} (Cambridge University Press,
  1996).

\bibitem{1990plos.book.....K}
D.~S. {Kliger}, J.~W. {Lewis}, and C.~E. {Randall}, \emph{{Polarized light in
  optics and spectroscopy}} (Academic Press, 1990).

\bibitem{1955PancharatnamAchromaticWaveplate}
S.~{Pancharatnam}, \enquote{{Achromatic combinations of birefringent plates},}
  Proc. Ind. Acad. Sci., A \textbf{41}, 137--144 (1955).

\bibitem{1950AbelesStratifiedMatrixMethod}
F.~{Abel\`{e}s}, \enquote{{Recherches sur la propagation des ondes
  \'{e}lectromagn\'{e}tiques sinusoidales dans les milieux stratifi\'{e}s},}
  Annales de Physique \textbf{5}, 596--640 (1950).

\bibitem{1999prop.book.....B}
M.~{Born} and E.~{Wolf}, \emph{{Principles of Optics}} (Cambridge University
  Press, 1999).

\bibitem{1987opt2.book.....H}
E.~{Hecht}, \emph{{Optics 2nd edition}} (Addison-Wesley Publishing Company,
  1987).

\bibitem{1972JOSA...62..502B}
D.~W. {Berreman}, \enquote{{Optics in Stratified and Anisotropic Media:
  4X4-Matrix Formulation},} Journal of the Optical Society of America
  (1917-1983) \textbf{62}, 502--510 (1972).

\bibitem{1979JOSA...69..742Y}
P.~{Yeh}, \enquote{{Electromagnetic propagation in birefringent layered
  media},} Journal of the Optical Society of America (1917-1983) \textbf{69},
  742--756 (1979).

\bibitem{EssingerHilemanPrincetonThesis}
T.~M. {Essinger-Hileman}, \enquote{{Probing Inflationary Cosmology: The Atacama
  B-Mode Search (ABS)},} Ph.D. thesis, Princeton University, New Jersey (2011).

\bibitem{2007A&A...470..771J}
W.~C. {Jones}, T.~E. {Montroy}, B.~P. {Crill}, C.~R. {Contaldi}, T.~S.
  {Kisner}, A.~E. {Lange}, C.~J. {MacTavish}, C.~B. {Netterfield}, and J.~E.
  {Ruhl}, \enquote{{Instrumental and analytic methods for bolometric
  polarimetry},} Astronomy \& Astrophysics \textbf{470}, 771--785 (2007).

\bibitem{2010ApOpt..49.6313B}
S.~A. {Bryan}, T.~E. {Montroy}, and J.~E. {Ruhl}, \enquote{{Modeling dielectric
  half-wave plates for cosmic microwave background polarimetry using a Mueller
  matrix formalism},} \ao \textbf{49}, 6313--+ (2010).

\bibitem{1983ChenCoordinateFreeApproach}
H.~{Chen}, \emph{{Theory of electromagnetic waves: A coordinate-free approach}}
  ({McGraw Hill, New York}, 1983).

\bibitem{1928HandbuchderPhysikSzivessy}
G.~{Szivessy}, \emph{{Handbuch der Physik}} (1928), vol.~20, chap.~11, p. 715.

\end{thebibliography}

\end{document}